\documentclass[12pt, oneside, onecolumn]{IEEEtran}
\usepackage{graphicx,dsfont,amssymb,amsmath,amsthm,cite,calc,tabulary,setspace}
\usepackage[caption=false,font=footnotesize]{subfig}
\usepackage{xcolor}
 \usepackage{algorithm,algpseudocode}
\usepackage{algpseudocode}
\usepackage{mathrsfs}

\usepackage{bbm}
\usepackage{color,soul}
\begin{document}
\title{Designing Asymmetric Shift Operators for Decentralized Subspace Projection}
\author{Siavash Mollaebrahim,~Baltasar~Beferull-Lozano,~\IEEEmembership{Senior Member,~IEEE}% <-this % stops a space
\thanks{The authors are with the Intelligent Signal Processing and Wireless Networks (WISENET) center, Department of Information and Communication Technology (ICT), University of Agder, Norway. e-mail:\{ siavash.mollaebrahim, baltasar.beferull\}@uia.no}
\thanks{This work was supported by the PETROMAKS Smart-Rig grant 244205/E30, the SFI Offshore Mechatronics grant 237896/O30, the IKTPLUSS INDURB grant 270730/O70 from the Research Council of Norway.}
}
\maketitle
\begin{abstract}
A large number of applications in wireless sensor networks includes projecting a vector of noisy observations onto a subspace dictated by prior information about the field being monitored. In general, accomplishing such a task in a centralized fashion, entails a large power consumption, congestion at certain nodes, and suffers from robustness issues against possible node failures. Computing such projections in a decentralized fashion is an alternative solution that solves these issues. Recent works have shown that this task can be done via the so-called graph filters where only local inter-node communication is performed in a distributed manner using a graph shift operator. Existing methods designed the graph filters for symmetric topologies. However, in this paper, the design of the graph shift operators to perform decentralized subspace projection for asymmetric topologies  is studied.%, which increases the feasible set of possible graph shift operators, leading additionally to an increase in performance with respect to the symmetric case.  
\end{abstract}
\section{Introduction}
Processing and analysis of data-sets gathered in different settings such as  social and economic networks, information networks, as well as infrastructure networks such as Wireless Sensor Networks (WSN) is of high importance in many applications where decentralized methods are required. 
These data-sets are structured and can be represented over graphs. Graph signal Processing (GSP) is a powerful tool that enables us to process and analyze graph-supported signals in different applications such as denoising, filtering and reconstruction of sensor data. %In this area, traditional signal processing processing tools defined on regular domains, are re-designed and developed to operate on the graph domain. For instance, the notations such as sampling and frequency analysis have been extended to the graph-supported signals~\cite{sandryhaila2013discrete, Shuman2013GSP, chen2015, sandryhaila2014}. 
In~\cite{sandryhaila2013discrete}, graph filters (GFs) have been introduced as polynomials of the so-called graph-shift operator, which is a local operator. Moreover, in~\cite{segarra2017operators}, the design of GFs to implement a pre-specific linear transformation has been studied.

In addition, it is important to notice that in general, processing data over WSN, Internet of Things (IoT) or other types of multi-agent networks, is one of the main goals of decentralized signal processing. %In many applications of interest, processing data should be performed in a decentralized fashion i.e only by exchanging local information among nodes. There are several reasons for this. For example, in centralized methods where there is fusion center, a node failure or the disruption of some communication link (either direct transmission or multi-hop communication) to the fusion center or sink node, supposes a clear limitation in terms of robustness with respect to a possible failure of the sink node or a node close to the sink node. Indeed, when a node failure or link disruption occurs for a node that is close to the sink node, this causes the loss of a large part of the sensor data. Furthermore, in some applications where there exist privacy issues, the agents are not allowed to send their local data to the sink node. Moreover, since in WSNs there are usually stringent power budget constraints, having continuous communication to a remote fusion center may also be unfeasible. On the other hand, decentralized processing enables us to perform in-network computations, allowing for instance some decisions to be made by intelligent nodes or agents. This property is vital for enabling real-time distributed wireless control in many industrial applications. 
Subspace projection is one of the most important instances of distributed data Processing. This problem not only can be regarded as a denoising or noise reduction task, but it is also directly connected to many other estimation problems \cite{barbarossa2009projection}. Let us assume a WSN network with a certain number of nodes $N$, where the gathered sensor measurements are assumed to be unreliable because of observation noise or erroneous data e.g. sensor malfunction, leading to a discrete noisy signal from which the original useful signal has to be estimated performing the subspace projection. In many physical fields of interest, the useful signal belongs typically to a subspace that has a dimension $r$ that is much smaller dimension than $N$. The subspace projection problem is to estimate the useful signal given the received noisy signal and the given subspace i.e. expressed in terms of a matrix whose columns span the useful signal subspace. 

%A decentralized subspace projection has been proposed in~\cite{barbarossa2009projection} based on a gossip algorithm. In this method, each node linearly combines its own iterate with its neighbours iterates, requiring an asymptotic convergence. Then, by optimizing a criterion that quantifies this asymptotic convergence, the coefficients of those linear combinations are found. However, this method needs a relatively high number of iterations to converge asymptotically. %Furthermore, it has been also shown that the proposed scheme in~\cite{barbarossa2009projection} can only be applied to a reduced set of symmetric topologies~\cite{camaro2013reducing}.    

GFs can be designed and implemented to perform decentralized subspace projection. In~\cite{segarra2017operators},~\cite{sandryhaila2014finitetime}, some GF-based methods have been proposed for a special case of subspace projection task, i.e. average consensus, which is a rank-1 projection. These methods are capable of converging after a finite number of iterations. In addition, more general scenarios of subspace projection have been considered in~\cite{segarra2017operators}. However, the proposed schemes need knowledge of the graph shift operator or matrix, which determines the linear combinations among the nodes. In most previous works, either the Laplacian matrix (or its normalized version) or the Adjacency matrix have been used as the default graph shift operator, or the design has been restricted to the case of consensus. To address these limitations, in~\cite{romero2018projection,mollaebrahim2018projection}, a symmetric graph shift operator has been designed to compute GFs that can accomplish decentralized subspace projection after a finite number of iterations.% Besides, in~\cite{Coutino2019Advance}, edge-variant GFs have been proposed to approximate linear transformations by solving a non-convex problem which is implementable for asymmetric topologies. Thus, their design does not allow for an exact implementation of a subspace projection operator. %The graph shift operator is obtained by optimizing a criterion that yields convergence to the subspace projection in a nearly minimal number of iterations. 

%It is important also to notice that a solution with an asymmetric graph shift is more suitable to be implemented over a WSN than a solution with a symmetric graph shift. Due to interferences and background noise, WSN are often characterized as time-varying asymmetric directed graphs. Most of existing works, however, assume symmetric GF designs to be implemented over symmetric communication topologies, which is not a realistic assumption in WSN.

The methods in~\cite{barbarossa2009projection,segarra2017operators,romero2018projection,mollaebrahim2018projection} have been proposed for undirected graph networks, which means that they consider symmetric topologies. However, in this paper, we consider the problem of decentralized subspace projection via GFs for asymmetric topologies. In addition, our design aims at making the filtering process efficient in terms of number of iterations. For this, we propose a new methodology based on the Schur matrix decomposition and formulate an optimization problem that exploits this decomposition, in order to obtain a GF that provides the exact subspace projection.% in contrast with the edge-variant GFs~\cite{Coutino2019Advance}, which can only approximate the projection matrix. %minimize the number of iterations (i.e. number of inter-node exchanges) until convergence. 

The contributions of this paper can be enumerated as follows: a) characterization of the existence and properties of a graph shift matrix, such that it is possible to construct a graph filter, exploiting the Schur Decomposition, and which implements the subspace projection operator, b) formulation of a convex optimization problem to design an efficient graph filter for an exact subspace projection considering also the required number of iterations, c) extensive experimental results showing the exact computation of the subspace projection and its superior performance as compared to current state-of-the-art methods. 

%The rest of the paper is structured as follows. In section II notations and reviews some existing results on decentralized subspace projection with graph filters is introduced. Section III presents the proposed algorithm. Finally, Section IV validates its performance through numerical experiments and section V concludes the paper.

\section{Decentralized Subspace Projection Problem}
Consider $N$ networked sensor nodes or agents that can exchange messages with their connected neighbors. The network is modeled as a directed connected graph $\mathcal{G}(\mathcal{V},\mathcal{E})$, where the vertex set $\mathcal{V}:={1,2,\cdots,N}$ correspond to the network agents, and $\mathcal{E}$ represents the set of edges. The $n'$-th vertex $v_{n'}$
is connected to $v_{n}$ if there is a directed edge
from $v_{n'}$ to $v_{n}$ $(v_{n'},{v}_{n}) \in \mathcal{E}$, but this does not mean that $v_{n}$ is
connected to $v_{n'}$ unless $(v_{n},{v}_{n'}) \in \mathcal{E}$. The in-neighborhood of the $n$-th node is defined as the set of nodes connected to it, which can be denoted as $\mathcal{N}_{n} = ({v_{n'}|(v_{n'},{v}_{n}) \in \mathcal{E}}$).

The observation vector $\mathbf{y}=\mathbf{x}+\mathbf{n}$ where $\mathbf{y}\in\mathbb{R}^{N}$ contains noisy information gathered by the nodes, where $\mathbf{x}$ and $\mathbf{n}$ are the useful signal and additive noise, respectively. The $n$-th entry of $\mathbf{y}=[y_1, y_2,\cdots,y_{N}]$ denotes information gathered by the $n$-th node. In the subspace projection context, the useful signal typically lies on a subspace of a dimension $r$ much smaller than $N$, which means that $\mathbf{x}=\mathbf{U}_{\parallel}\boldsymbol{\alpha}$ where $\mathbf{U}_{\parallel}\in\mathbb{R}^{N\times{r}}$ is a matrix whose columns span the useful signal subspace and $\boldsymbol{\alpha}\in\mathbb{R}^{r}$~\cite{barbarossa2009projection,romero2018projection}. 

Noise reduction can be obtained by projecting $\mathbf{y}$ onto the useful signal subspace which corresponds also to the least-squares estimate of $\mathbf{x}$, denoted by ${\hat{\mathbf{x}}}$, given by:  
\begin{align}
\hat{\mathbf{x}}=\mathbf{U}_{\parallel}\mathbf{U}_{\parallel}^{\top}\mathbf{y}\overset{\triangle}{=}\mathbf{P}\mathbf{y}
\end{align}
where $\mathbf{P}\in\mathbb{R}^{N\times{N}}$ is the projection matrix. Estimating the useful signal $\mathbf{x}$ from the observation signal $\mathbf{y}$ and the knowledge of the subspace matrix $\mathbf{U}_{\parallel}$, is the subspace projection problem. 
 
%A decentralized scheme for subspace projection has been proposed in~\cite{barbarossa2009projection}, which is based on a iterative approach. The iterates $\mathbf{z}[k+1]=\mathbf{W}\mathbf{z}[k]\quad\forall{k=0,\cdots,N}$ with initialization $\mathbf{z}[0]=\mathbf{z}$. This scheme tries to find a sparse matrix $\mathbf{W}$ such that i) $(\mathbf{W})_{n,n'}=0$ if $(v_n,v_n')\notin\mathcal{H}$, ii) $\lim_{k\to\infty}\mathbf{z}[k]=\lim_{k\to\infty}\mathbf{W}^{k}\mathbf{z}=\mathbf{P}\mathbf{z}, \forall{\mathbf{z}\in}\mathbb{R}^{N}$. The convergence of this method is asymptotic, and it needs a large number of iterations to converge asymptotically. Also, the set of feasible topologies that can satisfy the constraints (i), (ii) is limited~\cite{camaro2013reducing}. Moreover, this approach has been restricted to the symmetric matrices $\mathbf{W}$. 

As stated earlier, decentralized subspace projection can be performed by using GFs. In this case, we are also given a certain asymmetric network connectivity topology so that each node has a certain set of neighbour nodes it can reach and also there is a certain set of neighbour nodes from which it can be reached. Let us first introduce the concept of graph shift matrix. Any matrix that satisfies $\mathbf{S}_{(n,n')}=0$ if $(v_n,v_n')\notin\mathcal{E}$ is a feasible graph shift matrix, which implies also that $\mathbf{S}$ characterises the underlying network topology~\cite{sandryhaila2013discrete}. 
% COMMENT: This part is not relevant for the conference paper
%An attractive feature of a graph shift matrix $\mathbf{y}\mapsto{\mathbf{S}\mathbf{y}}$ is that it can be computed in a decentralized fashion. Examples of $\mathbf{S}$ include the adjacency matrix or the Laplacian matrix. 

A graph filter is a linear combination of successively shifted graph signals i.e. $\mathbf{H}:=\sum_{l=0}^{L-1}c_l\mathbf{S}^{l}$ where $\{c_l\}_{l=0}^{L-1}$ are the filter coefficients, and $L$ is the order of the filter. The procedure is that all nodes exchange their information with their neighbours ($y_n$ is the $n$-th node observed noisy signal sample), so that the signal information of all nodes is updated via $\mathbf{y}^{(1)}=\mathbf{S}\mathbf{y}$. For the next iteration, we have $\mathbf{y}^{(2)}=\mathbf{S}\mathbf{y}^{(1)}=\mathbf{S}^{2}\mathbf{y}$. This procedure is repeated for $L$ iterations. Thus, a graph filter can be computed in a decentralized fashion. To compute the subspace projection via GFs, it is needed that $\mathbf{P}=\mathbf{H}$. It has been shown previously in~\cite{segarra2017operators} that with some appropriate given choices of the shift matrix (typically Laplacian or Adjacency matrices), the filter coefficients for $\mathbf{P}=\sum_{l=0}^{L}c_l\mathbf{S}^{l}$ can be found by solving a linear system of equations. GFs have been also designed for rank-1 projections in~\cite{segarra2017operators}, which is a special case of subspace projection.  In~\cite{romero2018projection}, valid symmetric shift matrices are found by optimizing a criterion related to minimizing the filter order ($L$), i.e. the number of iterations or node information exchanges required for convergence to the projection matrix after a finite number of iterations. In this paper, 
we consider the formulation of an optimization problem aiming at finding an asymmetric graph shift matrix $\mathbf{S}$ that computes the exact subspace projection. %minimizes the required filter order, that is, removing the constraint of symmetry in the graph shift operator. 
%As we show next, allowing for asymmetric matrices, makes it necessary to develop a different design method. 
%In all these works, it is necessary to have knowledge of the graph shift matrix.
%Moreover, for the scenarios beyond rank-1 projections, knowledge of the shift matrix is %needed which is seldom known. Therefore, finding a valid $\mathbf{S}$ is the key %challenge. To answer this challenge,   
\section{Problem Formulation and Proposed method}
Our main problem can be formally stated as follows: 

\textbf{Given}: i) A matrix $\mathbf{U}_{\parallel}  \in \mathbb{R}^{N \times r}$ whose columns span the subspace of interest and ii) the set of edges $\mathcal {E}=\{(n_1,n_{1}^{'}), (n_2,n_{2}^{'}), \ldots, (n_{|\mathcal {E}|},n_{|\mathcal {E}|}^{'})\}$ defining an asymmetric network topology.

\textbf{Find}: An asymmetric matrix $\mathbf{S}\in{\mathbb{R}}^{N\times{N}}$ (the graph shift matrix) and a vector $\mathbf{c}=[c_0, c_1,...,c_{L-1}]^{\top}\in{\mathbb{R}}^{L}$ (the filter coefficients) \textbf{such that}: 

\begin{itemize}
\item $\sum_{l=0}^{L-1}c_l\mathbf{S}^{l}=\mathbf{U}_{\parallel}\mathbf{U}_{\parallel}^{\top}\Rightarrow$ Polynomial shift matrix

\item $\mathbf{S}_{n,n'}=0$ $\forall{(v_n,v_{n'})}\not\in{\mathcal{E}}$ $\Rightarrow$ Topological shift matrix
\end{itemize}
The first and the second condition mean that $\mathbf{S}$ is polynomially and topologically feasible, respectively. Our goal in this paper is to find shift matrices that are both polynomial and topological. 
In order to achieve this, we need to introduce the following Schur matrix factorization.% Note that without loss of generality, in this paper, $\mathbf{S}$ is a real matrix with real eigenvalues. %let us consider a matrix factorization method called as the Schur decomposition.
%\textit{\textbf{Theorem 1} (Schur decomposition)}:

\textit{\textbf{Schur Decomposition}}: If $\mathbf{S}\in{\mathbb{R}}^{N\times{N}}$, then it can be decomposed as follows~\cite{Golub}:
\begin{align}
\mathbf{S}=\mathbf{W}\mathbf{T}\mathbf{W}^{T}
\end{align}
 where $\mathbf{T}\in{\mathbb{R}}^{N\times{N}}$ is upper quasi-triangular and $\mathbf{W}\in{\mathbb{R}}^{N\times{N}}$ is orthogonal. In fact, $\mathbf{T}$ has the following form:
\begin{align}
\begin {bmatrix} \mathbf{B}_{1}&*&\cdots&*\\\mathbf{0}&\mathbf{B}_{2}&\cdots&*\\\vdots&\vdots&\ddots&\vdots\\\mathbf{0}&\mathbf{0}&\cdots&\mathbf{B}_{k}\end {bmatrix}
\end{align}
 The diagonal blocks $\mathbf{B}_i$ are either $1\times{1}$ or $2\times{2}$ matrices.  A $1 \times 1$ block corresponds to a real eigenvalue, and a $2 \times 2$ block has four real entries that correspond to the real and imaginary parts of a pair of complex conjugate eigenvalues of $\mathbf{S}$. 
%Even though it is possible to provide the results of this paper to asymmetric matrices $\mathbf{S}$ with complex eigenvalues, 
In this paper, for the sake of simplicity in the proof of our main results, we restrict our attention to asymmetric matrices $\mathbf{S}$ with real eigenvalues. 
  %Therefore, we can conclude that the result of \textbf{Theorem 1} holds for real matrices whose all eigenvalues are real. %In this paper, we aim to find such matrices. 
For an asymmetric matrix $\mathbf{S}$ with real eigenvalues, 
the following decomposition holds:
%for the sake of simplicity, we restrict our attention to asymmetric matrices $\mathbf{S}$ with real %eigenvalues. Consequently, an asymmetric matrix $\mathbf{S}$ can be decomposed as:
\begin{align}\label{shift}
\mathbf{S}=\mathbf{W}(\mathbf{D}+\mathbf{Q})\mathbf{W}^{\top}
\end{align}
where $\mathbf{D}$ is diagonal, $\mathbf{Q}$ is upper triangular with diagonal elements being zero and $\mathbf{W}$ is orthogonal. %The next Theorem states that the problem of finding  polynomial asymmetric shift matrices is feasible.  % In our case, by using schur decomposition, we have $\sum_{l=0}^{L}c_l\mathbf{S}^{l}=\sum_{l=0}^{L}c_l\mathbf{W}(\mathbf{D}+\mathbf{Q})^{l}\mathbf{W}^{\top}$. 
%From the binomial theorem, we have:
%\begin{align}
%\sum_{l=0}^{L}c_l\mathbf{W}(\mathbf{D}+\mathbf{Q})^{l}\mathbf{W}^{\top}=\sum_{l=0}^{L}c_l\mathbf{W}(\mathbf{D})^{l}\mathbf{W}^{\top}+\sum_{l=0}^{L}c_l(\sum_{k=1}^{l}{l\choose{k}}\mathbf{W}(\mathbf{D})^{l-k}\mathbf{Q}^{k}\mathbf{W}^{\top})
%\end{align}
 %Consequently, $\sum_{l=0}^{L}c_l\mathbf{W}(\mathbf{D}+\mathbf{Q})^{l}\mathbf{W}^{\top}\approx\mathbf{U}_{\parallel}\mathbf{U}_{\parallel}^{\top}$ can be satisfied by:
%\begin{align}
%\sum_{l=0}^{L}c_l\mathbf{W}(\mathbf{D})^{l}\mathbf{W}^{\top}=\mathbf{U}_{\parallel}\mathbf{U}_{\parallel}^{\top}\label{asym1}\\
%\sum_{l=0}^{L}c_l(\sum_{k=1}^{l}{l\choose{k}}\mathbf{W}(\mathbf{D})^{l-k}\mathbf{Q}^{k}\mathbf{W}^{\top})\approx\mathbf{0}\label{asym2}
%\end{align} 
%\section{Designing the graph shift operator}
Let us consider $\mathbf{D}=\left(	\begin{matrix} \mathbf{D}_{1}& \mathbf{0}\\ \mathbf{0}&\mathbf{D}_{2}\end{matrix} \right)$,
$\mathbf{D}_1 = \textbf{diag}([\lambda_1, \cdots, \lambda_{r}])$ and $\mathbf{D}_2 = \textbf{diag}([\lambda_{r+1}, \cdots,\lambda_N])$ with all the diagonal elements being real values.

\textit{\textbf{Proposition 1}}:\textit{ Given a matrix $\mathbf{U}_{\perp}\in{\mathbb{R}}^{N\times{N-r}}$ with orthonormal columns that satisfy ${\mathcal{R}(\mathbf{U}_{\perp})=\mathcal{R}^{\perp}(\mathbf{U}_{\parallel})}$, $\mathbf{P}=\mathbf{U}_{\parallel}\mathbf{U}_{\parallel}^{\top}$, a necessary condition to have a polynomial graph shift operator is that $\mathbf{D}_{1}$ and $\mathbf{D}_{2}$ do not share any eigenvalue}.
\vspace{-2mm}
\begin{proof}
%then there exists $\{c_{l}\}$ and an asymmetric polynomial shift matrix $\mathbf{S}\in\mathbb{R}^{N\times{N}}$ such that $\sum_{l=0}^{L-1} c_l \mathbf{S}^{l}=\mathbf{P}$ with $L={N}$
%\section{Discussion}
%In this section, we provide necessary and sufficient conditions, which guaranty that there exist an asymmetric $\mathbf{S}\in\mathbb{R}^{N\times{N}}$ and $\{c_{l}\}$ such that $\sum_{l=0}^{L}c_l\mathbf{S}^{l}=\mathbf{P}$.  %As we show next, it is sufficient to consider the set of matrices $\mathbf{S}$ with real eigenvalues for finding a valid polynomial shift matrix. 

From the Schur decomposition of $\mathbf{S}$, %we have that  $\sum_{l=0}^{L}c_l\mathbf{S}^{l}=\sum_{l=0}^{L}c_l\mathbf{W}(\mathbf{D}+\mathbf{Q})^{l}\mathbf{W}^{\top}$
and the condition for $\mathbf{S}$ to be a polynomial shift matrix, we have:
\begin{align}\label{schur0}
    &\sum_{l=0}^{L-1}c_l\mathbf{S}^{l}=\sum_{l=0}^{L-1}c_l[\mathbf{W}_{\parallel}\quad\mathbf{W}_{\perp}](\mathbf{D}+\mathbf{Q})^{l}[\mathbf{W}_{\parallel}\quad\mathbf{W}_{\perp}]^{\top}=\nonumber\\&[\mathbf{U}_{\parallel}\quad\mathbf{U}_{\perp}]\left(	\begin{matrix} \mathbf{I}_{r}& \mathbf{0}\\ \mathbf{0}&\mathbf{0}\end{matrix} \right)[\mathbf{U}_\parallel\quad\mathbf{U}_{\perp}]^{\top}
\end{align}
%\noindent where: %${\mathcal{R}(\mathbf{U}_{\perp})=\mathcal{R}^{\perp}(\mathbf{U}_\parallel)}$. 
\noindent{where} $\mathbf{W}=\begin {bmatrix} \mathbf{W}_{\parallel}& \mathbf{W}_{\perp}\end {bmatrix}$. % and $\mathbf{W}_{\parallel}\in{\mathbb{R}^{N\times{r}}}, \mathbf{W}_{\perp}\in{\mathbb{R}^{N\times{N-r}}}$. 
 From \eqref{schur0}, if $\exists \{\mathbf{c},\mathbf{D}, \mathbf{Q}\}:\sum_{l=0}^{L-1}c_l(\mathbf{D}+\mathbf{Q})^{l}=\left(	\begin{matrix} \mathbf{I}_{r}& \mathbf{0}\\ \mathbf{0}&\mathbf{0}\end{matrix} \right)$, then $\mathbf{W}_{\parallel}\mathbf{W}_{\parallel}^{\top}=\mathbf{U}_\parallel\mathbf{U}^{\top}_\parallel$ (implies that $\mathcal{R}(\mathbf{W}_{\parallel})=\mathcal{R}(\mathbf{U}_\parallel)$), and there exists a feasible polynomial shift matrix $\mathbf{S}$.% where $\mathbf{c}=[c_{0},c_{1},\cdots, c_{N-1}]^{\top}$. %Then, we have that:
%\begin{align}\label{schur1}
%\sum_{l=0}^{L}c_l\mathbf{S}^{l} = \sum_{l=0}^{L}c_l\mathbf{W}(\mathbf{D}+\mathbf{Q})^{l}\mathbf{W}^{\top}=\mathbf{W}_{\parallel}\mathbf{W}_{\parallel}^{\top}=\mathbf{U}_\parallel\mathbf{U}^{\top}_\parallel=\mathbf{P}
%\end{align}
%which implies that $\mathbf{S}$ is a polynomial shift matrix. 

The required $\{\mathbf{c}, \mathbf{D}, \mathbf{Q}\}$ should satisfy:
%We need to have $\{\mathbf{c}, \mathbf{D}, \mathbf{Q}\}$ such that:
\begin{align}\label{asim1}
\sum_{l=0}^{L-1}c_l(\mathbf{D}+\mathbf{Q})^{l}=\left(	\begin{matrix} \mathbf{I}_{r}& \mathbf{0}\\ \mathbf{0}&\mathbf{0}\end{matrix} \right)
\end{align}
%If we have $L={(N^2+N)/2}$, then \eqref{asim1} holds.
%To find the filter coefficient vector ($\mathbf{c}= [c_0, c_1, \ldots, c_{L-1}]^T$), 
%\begin {bmatrix} c_{0}\\ c_{1}\\\vdots\\c_ {L-1}\end {bmatrix}$  ), 
We can rewrite \eqref{asim1} as a set of equations in the form of $\mathbf{T}\mathbf{c}=\mathbf{b}$, where $\mathbf{c}\in\mathbb{R}^{L\times{1}}$ and $\mathbf{T}\in\mathbb{R}^{N^{2}\times{L}}$ and  $\mathbf{T}=(\mathrm{vec}(\mathbf{Z}^{0})| \mathrm{vec}(\mathbf{Z})|\cdots|\mathrm{vec}(\mathbf{Z}^{L}))$, $\mathbf{Z}=\mathbf{D}+\mathbf{Q}$ and $\mathbf{b}=\mathrm{vec}\left(	\begin{matrix} \mathbf{I}_{r}& \mathbf{0}\\ \mathbf{0}&\mathbf{0}\end{matrix} \right)$. 
Since $\mathbf{D}$ is diagonal and $\mathbf{Q}$ is upper triangular, $\mathbf{D} + \mathbf{Q}$ is upper triangular, and then $(N^2-N)/2$ rows of $\mathbf{T}$ are zero. Thus, we can remove them reducing the size of the matrix $\mathbf{T}$ to another matrix $\mathbf{T'}\in\mathbb{R}^{N_{1}\times{L}}$ where $N_{1}=((N^2+N)/2)$. 
%This is the case when all diagonal elements of %$\mathbf{D}$ are distinct. 

 % and we show that it is possible to find a valid solution in this case. 
%$\mathbf{D}=\begin {bmatrix} \lambda_{1}&0&\cdots&0\\ 0&\lambda_{2}&\cdots&0\\\vdots&0&\ddots&0\\0&0&\cdots&\lambda_{N}\end {bmatrix}$
  Moreover, by direct computation in \eqref{asim1} using the Binomial theorem, some rows of $\mathbf{T'}$ consist of powers of eigenvalues in $\mathbf{S}$ (this corresponds to powers of matrix $\mathbf{D}$), and the other ones correspond to products of matrices $\mathbf{Q}$ and $\mathbf{D}$ with diagonal entries being zero. Let us consider in \eqref{asim1} only the $N$ rows of $\mathbf{T'}$ associated to the powers of $\mathbf{D}$, leading to the following set of equations:    
\begin{align}\label{rowasim}
\begin{bmatrix}1&\lambda_1&\lambda_1^2&\cdots&\lambda_1^L\\{\vdots}&\vdots&\vdots&\vdots&\vdots\\1&\lambda_r&\lambda_r^2&...&\lambda_r^L\\{\vdots}&\vdots&\vdots&\vdots&\vdots\\1&\lambda_N&\lambda_N^2&\cdots&\lambda_N^L\\\end{bmatrix}\begin{bmatrix}c_1\\c_2\\\vdots\\c_L\\\end{bmatrix}=\begin{bmatrix}\mathbf{1}_r\\\mathbf{0}_{N-r}\\\end{bmatrix}
\end{align}
%If $L={(N^2+N)/2}$ and the columns of $\mathbf{T'}$ are independent, the filter coefficients can be found easily since in this case, $\mathbf{T'}$ is square and full rank, which means that it is also invertible. 
%which is a Vandermonde matrix. 
From \eqref{rowasim}, the necessary condition to have a solution in \eqref{asim1} is that $\mathbf{D}_1$ and $\mathbf{D}_2$ do not share any eigenvalue.
\end{proof}
The sufficient condition to have a solution or solutions in \eqref{asim1} is that $L\ge{N_{1}-m_{1}-m_{2}}$ and $\mathbf{T'}$ should be a full-row rank matrix where $m_{1}$ and $m_{2}$ are the number of eigenvalues with multiplicity larger than one in $\mathbf{D}_{1}$ and $\mathbf{D}_{2}$, respectively. In this case, there exists $\mathbf{S}$ that satisfies $\sum_{l=0}^{L-1}c_l\mathbf{S}^{l}=\mathbf{P}$ with $L\ge{N_{1}-m_{1}-m_{2}}$. Then, based on the Cayley-Hamilton theorem, $\exists{\hat{\mathbf{c}}}$ such that  $\sum_{l=0}^{L}\hat{c}_l\mathbf{S}^{l}=\mathbf{P}$ with $L\le{N-1}$. Please note that in \eqref{rowasim}, if there are some repeating eigenvalues in $\mathbf{D}_{1}$ or $\mathbf{D}_{2}$, we can remove the rows associated with them in  equation \eqref{rowasim}.

Next, we propose a method to obtain a shift matrix that is both polynomial and topological $\mathbf{S}\in\mathbb{R}^{N\times{N}}$ by formulating a convex optimization problem, 
enforcing an optimization criterion that promotes an efficient filtering, leading to a unique solution.
First of all, based on the proof of \textit{\textbf{Proposition 1}}, we can take  $\mathbf{W}_{\parallel}=\mathbf{E}_{\parallel}\mathbf{U}_{\parallel}$ and $\mathbf{W}_{\perp}=\mathbf{E}_{\perp}\mathbf{U}_{\perp}$ where $\mathbf{E}_\parallel, \mathbf{E}_{\perp}$ are arbitrary orthogonal matrices. %to have a polynomial shift matrix, we need to have $\sum_{l=0}^{L}c_l\mathbf{S}^{l}=\mathbf{U}_{\parallel}\mathbf{U}_{\parallel}^{\top}$. If $\mathbf{S}$ is decomposed based on its Schur decomposition, from the polynomial condition \eqref{asim1}, we have:
 %\begin{align}\label{sum}
  %   \sum_{l=0}^{L}c_l\mathbf{S}^{l}=\sum_{l=0}^{L}c_l\mathbf{W}(\mathbf{D}+\mathbf{Q})^{l}\mathbf{W}^{\top}=\mathbf{U}_{\parallel}\mathbf{U}_{\parallel}^{\top}
 %\end{align}
 %\eqref{sum} can be rewritten as:
 %\begin{align}\label{sum2}
  %    &\sum_{l=0}^{L}c_l[\mathbf{W}_{\parallel}\quad\mathbf{W}_{\perp}](\mathbf{D}+\mathbf{Q})^{l}[\mathbf{W}_{\parallel}\quad\mathbf{W}_{\perp}]^{\top}=[\mathbf{U}_{\parallel}\quad\mathbf{U}_{\perp}]\left(	\begin{matrix} \mathbf{I}_{r}& \mathbf{0}\\ \mathbf{0}&\mathbf{0}\end{matrix} \right)[\mathbf{U}_\parallel\quad\mathbf{U}_{\perp}]^{\top}
 %\end{align}
 %\noindent where: ${\mathcal{R}(\mathbf{U}_{\perp})=\mathcal{R}^{\perp}(\mathbf{U}_{\parallel})}$. Based on the proof of \textbf{proposition 1}, we need to have: $\mathbf{W}_{\parallel}\mathbf{W}_{\parallel}^{\top}=\mathbf{U}_\parallel\mathbf{U}^{\top}_\parallel$, with the elements of $\mathbf{D}$ being distinct.  %, finding a polynomial shift matrix is feasible.  
 %Thus, to design . This guaranties that $\mathbf{W}_{\parallel}\mathbf{W}_{\parallel}^{\top}=\mathbf{U}_{\parallel}\mathbf{U}_{\parallel}^{\top}$. 
 Therefore, in the next steps, we only need to focus on the design of the matrices $\mathbf{D}$ and $\mathbf{Q}$. %For the rest of paper, we assume, without loss of generality, that the matrix $\mathbf{S}$ has real eigenvalues. 

We propose to use the cost function $||\mathbf{Q}||_{F}^{2}+{{\left\| \mathbf{D}_{2}  \right\|}_{F}^{2}}$ in order to both increase numerical stability when obtaining the filter coefficients and also promote a fast convergence to the subspace projection. This cost function is also well-motivated when the links between nodes are noisy and finding a low energy shift matrix is desirable. Moreover, from the right-hand side of \eqref{asim1}, we can observe that the upper right hand-side block matrix is a zero matrix; since we know from the Schur decomposition of $\mathbf{S}$ that $\mathbf{D}$ and $\mathbf{Q}$ are diagonal and upper triangular, respectively, then, minimizing $\mathbf{Q}$ and $\mathbf{D}_2$ in Frobenius norm will also promote faster graph filters. %In fact, after applying the filter coefficients, the terms associated with $\mathbf{D}_{2}, \mathbf{Q}$ should become zero. Minimizing $||\mathbf{D}_{2}||_{F}^{2}+||\mathbf{Q}||_{F}^{2}$ helps us to reach that goal, and decreasing the effect of additive noise. 

Consequently, our optimization problem is stated as follows:
%to prevent any numerical difficulties during obtaining the filter coefficients especially at lower orders of the filter ($l$), we decrease the effect of $\mathbf{Q}$ which helps us to make the convergence of the filter faster. %we can conclude that if Q gets sparse, then the order of our filter can decrease (note that, for instance if $\mathbf{Q}=\mathbf{0}$ then $L=N$). 
\begin{subequations} \label{oneproblem} 
\begin {align}
  \underset{\mathbf{S},\mathbf{D}_{1},\mathbf{D}_{2},\mathbf{Q}}{\text{min}} &
    %    \lambda{{\left\| \mathbf{U}_{\parallel}\mathbf{F}_{\parallel}\mathbf{U}_{\parallel}^{\top} + \mathbf{U}_{\perp}\mathbf{F}_{\perp}\mathbf{U}_{\perp}^{\top}\right\|}_{F}^{2}}+
{{\left\| \mathbf{Q}  \right\|}_{F}^{2}}+{{\left\| \mathbf{D}_{2}  \right\|}_{F}^{2}}\\
\text{s. t.}   \;\;\;\;    	&  (\mathbf S)_{n,n'}=0 \hspace{2mm} \text{if} \hspace{2mm} (v_n,v_{n'})\not \in \mathcal{E} , \; n,n' =1,\cdots,N  \label{topo}\\
		%	& \mathbf {F}_{\parallel}=\mathbf {F}_{\parallel}^{\top}, \mathbf {F}_{\perp}=\mathbf {F}_{\perp}^{\top}\\
  &  \mathbf{S}=\mathbf{W}_{\parallel}\mathbf{D}_{1}\mathbf{W}_{\parallel}^{\top}+\mathbf{W}_{\perp}\mathbf{D}_{2}\mathbf{W}_{\perp}^{\top}+\mathbf{W}\mathbf{Q}\mathbf{W}^{\top}\\
&  (\mathbf Q)_{l,l'}=0 , \; l'\le{l}, \;\quad l,l'=1,\cdots,N\label{consasym}\\
&  \mathbf{X}_{1}\mathrm{vec}(\mathbf{D})=\mathbf{0}, \; \mathbf{X}_{2}\mathrm{vec}(\mathbf{D})=\mathbf{0}\label{consasym11}\\
%&\boldsymbol{\lambda_1}(i)\neq{\boldsymbol{\lambda_2}(j)}%,\quad \forall{i,j} \label{eig}\\
%&\boldsymbol{\lambda_1} (i)\neq \boldsymbol{\lambda_1}(j) %\quad \boldsymbol{\lambda_2} (i)\neq %\boldsymbol{\lambda_2}(j) \quad \forall{i,j}, i\neq{j} 
&\lambda_{1i}  \neq  \lambda_{2j},\quad \forall{i,j} \label{eig}
%&\lambda_{1i}  \neq  \lambda_{1j},\quad  \lambda_{2i} \neq \lambda_{2j}, \quad \forall{i,j}, \; i \neq j
%\label{eig2}
\end{align}
\end{subequations}
\noindent{where} $\boldsymbol{\lambda_1} = [\lambda_{11}, \cdots, \lambda_{1r}]$, $\boldsymbol{\lambda_2}= [\lambda_{21}, \cdots, \lambda_{2 (N-r)}]$ are the vectors containing the real eigenvalues of $\mathbf{D}_{1}$ and $\mathbf{D}_{2}$, respectively, where we have renamed the eigenvalues in \eqref{rowasim}. Besides, \eqref{consasym11} makes $\mathbf{D}_{1}$, $\mathbf{D}_{2}$ diagonal (i.e. matrices $\mathbf{X}_{1}$ and $\mathbf{X}_{2}$ select non-diagonal elements of $\mathbf{D}_{1}$ and $\mathbf{D}_{2}$, respectively) and \eqref{eig} is added to the problem because based on \textit{\textbf{proposition 1}} the eigenvalues of $\mathbf{D}_{1},\mathbf{D}_{2}$ should be distinct. Notice also that, as explained before, the matrix $\mathbf{W}$ can be chosen in advance and is not part of the problem. In addition, since $\mathbf{S}$ is a local operator, \eqref{topo} is also added to enforce the asymmetric topological constraints.  
However, optimization problem \eqref{oneproblem} is non-convex because of \eqref{eig}. In addition, $\mathbf{S}=\mathbf{0}$ is one of its solutions which is a trivial solution. In order to tackle this problem and make the eigenvalues of $\mathbf{D}_1$ and $\mathbf{D}_2$ distinct, we can restrict the values of $\mathbf{D}_{1}$ by adding $\mathrm{tr}(\mathbf{D}_{1})=\epsilon{r}$ as a new term to the problem where $\epsilon$ can be chosen arbitrarily. 

Thus, we have:
\begin{subequations} \label{final} 
\begin {align}
  \underset{\mathbf{S},\mathbf{D}_{1},\mathbf{D}_{2},\mathbf{Q}}{\text{min}} &
    %    \lambda{{\left\| \mathbf{U}_{\parallel}\mathbf{F}_{\parallel}\mathbf{U}_{\parallel}^{\top} + \mathbf{U}_{\perp}\mathbf{F}_{\perp}\mathbf{U}_{\perp}^{\top}\right\|}_{F}^{2}}+
{{\left\| \mathbf{Q}  \right\|}_{F}^{2}}+{{\left\| \mathbf{D}_{2}  \right\|}_{F}^{2}}\\
\text{s. t.}   \;\;\;\;    	&  (\mathbf S)_{n,n'}=0 \hspace{2mm} \text{if} \hspace{2mm} (v_n,v_{n'})\not \in \mathcal{E} ,  n,n' =1,\cdots,N \label{topological} \\
		%	& \mathbf {F}_{\parallel}=\mathbf {F}_{\parallel}^{\top}, \mathbf {F}_{\perp}=\mathbf {F}_{\perp}^{\top}\\
&\mathrm{tr}(\mathbf{D}_{1})=r\epsilon\label{tracf}\\
  &  \mathbf{S}=\mathbf{W}_{\parallel}\mathbf{D}_{1}\mathbf{W}_{\parallel}^{\top}+\mathbf{W}_{\perp}\mathbf{D}_{2}\mathbf{W}_{\perp}^{\top}+\mathbf{W}\mathbf{Q}\mathbf{W}^{\top}\label{shift}\\
&  (\mathbf Q)_{l,l'}=0 , l'\le{l},\quad l,l'=1,\cdots,N\label{consasym1}\\
&  \mathbf{X}_{1}\mathrm{vec}(\mathbf{D})=\mathbf{0}, \mathbf{X}_{2}\mathrm{vec}(\mathbf{D})=\mathbf{0}\label{consasym6}
\end{align}
\end{subequations}
Please note that, if the rows associated with the combinations of $\mathbf{D}$ and $\mathbf{Q}$ make $\mathbf{T'}$ rank deficient, then our proposed method approximates the projection matrix. Although, based on our numerical results, this never happened. To deal with this issue, after obtaining $\mathbf{S}$ via the optimization problem, rank $(\mathbf{T'})$ is checked. If the matrix is not full rank, we solve the optimization problem with different $\epsilon$ to achieve a full-rank $\mathbf{T'}$. As we show next, optimization problem \eqref{final} can be solved via the Alternating Direction Method of Multipliers (ADMM)~\cite{Boyd2011ADMM} effectively. For this, by substituting \eqref{shift} into \eqref{topological}, the optimization problem \eqref{final} can be rewritten based on $\mathbf{D}$ and $\mathbf{Q}$, as follows: 
\begin{subequations} \label{final2} 
\begin {align}
  \underset{\mathbf{D},\mathbf{Q}}{\text{min}} &
    %    \lambda{{\left\| \mathbf{U}_{\parallel}\mathbf{F}_{\parallel}\mathbf{U}_{\parallel}^{\top} + \mathbf{U}_{\perp}\mathbf{F}_{\perp}\mathbf{U}_{\perp}^{\top}\right\|}_{F}^{2}}+
{{\left\| \mathrm{vec}({\mathbf{Q}})  \right\|}_{2}^{2}}+{{\left\| \mathbf{F}\mathrm{vec}({\mathbf{D}})  \right\|}_{2}^{2}}\\
\text{s. t.}   \;\;\;\;    	&  \mathbf{T}\mathrm{vec}(\mathbf{W}\mathbf{D}\mathbf{W}^{\top}+\mathbf{W}\mathbf{Q}\mathbf{W}^{\top})=\mathbf{T}(\mathbf{W}\otimes\mathbf{W})\nonumber\\&\mathrm{vec}(\mathbf{D}+\mathbf{Q}) =\mathbf{0} \\
		%	& \mathbf {F}_{\parallel}=\mathbf {F}_{\parallel}^{\top}, \mathbf {F}_{\perp}=\mathbf {F}_{\perp}^{\top}\\
&  \mathbf{P}\mathrm{vec}(\mathbf D)=\mathbf{b} , \mathbf{R}\mathrm{vec}(\mathbf Q)=\mathbf{0}
\end{align}
\end{subequations}
\noindent{where} we have the following matrices: $\mathbf{F}$ is a matrix that satisfies $\mathbf{F}\mathrm{vec}(\mathbf{D})=\mathrm{vec}(\mathbf{D}_2)$,  $\mathbf{T}$ has a row $(\mathbf{e}_{n}\otimes\mathbf{e}_{n'})^{\top}$ for each pair $(n,n')$ such that ${(v_n,v_{n'})}\not\in{\mathcal{E}}$, and $\mathbf{e}_{n}$ represents the $n$-th column of the identity matrix with the corresponding size. 
We can write \eqref{tracf} in vector form as $\mathbf{y}\mathrm{vec}(\mathbf{D})=r{\epsilon}$ where $\mathbf{y}$ is a vector whose entries are zero and one such that satisfies \eqref{tracf}. Similarly, we can find $\mathbf{R}$ such that $\mathbf{R}\mathrm{vec}(\mathbf{Q})=\mathbf{0}$ which satisfies~\eqref{consasym1}. Therefore, we have $\mathbf{P}=[\mathbf{y};\mathbf{X}_1;\mathbf{X}_2], \mathbf{b}=[r{\epsilon},\mathbf{0},\mathbf{0}]$.

Notice that we have now an optimization problem with a convex cost function and linear constraints, making it possible to apply the scaled-form of ADMM to solve this problem efficiently. We have that:
  \begin{align}
   &L_{\rho}(\mathbf{D},\mathbf{Q},\mathbf{v}_{1},\mathbf{v}_{2}, \mathbf{v}_{3})=(\rho/2) {{\left\|\mathbf{M}\mathrm{vec}(\mathbf{D}+\mathbf{Q})+\mathbf{v}_{1}\right\|}_{2}^{2}}\nonumber\\&+{{\left\| \mathrm{vec}(\mathbf{Q}) \right\|}_{2}^{2}}+(\rho/2){{\left\| \mathbf{P}\mathrm{vec}(\mathbf{D})-\mathbf{b}+\mathbf{v}_{2}\right\|}_{2}^{2}}\nonumber\\&+(\rho/2){{\left\| \mathbf{R}\mathrm{vec}(\mathbf{Q})+\mathbf{v}_{3}\right\|}_{2}^{2}}+{{\left\| \mathbf{F}\mathrm{vec}(\mathbf{D})\right\|}_{2}^{2}}
  \end{align}
  
  \noindent{where} $\mathbf{M}=\mathbf{T}(\mathbf{W}\otimes\mathbf{W})$. The closed form solutions for each ADMM iteration is given as follows:
  \begin{align}\label{fin1}
   &\mathrm{vec}(\mathbf{Q}[k+1])=-(\rho\mathbf{M}^{\top}\mathbf{M}+\mathbf{I}+\rho\mathbf{R}^{\top}\mathbf{R})^{-1}\nonumber\\&\rho(\mathbf{M}^{\top}(\mathrm{vec}(\mathbf{D}[k])+\mathbf{v}_1[k])+\mathbf{R}^{\top}\mathbf{v}_3[k])
  \end{align}
  \begin{align}\label{fin2}
  & \mathrm{vec}(\mathbf{D}[k+1])=-(\rho\mathbf{M}^{\top}\mathbf{M}+\rho\mathbf{P}^{\top}\mathbf{P}+\mathbf{F}^{\top}\mathbf{F})^{-1}\rho(\mathbf{M}^{\top}\nonumber\\&(\mathrm{vec}(\mathbf{Q}[k+1])+\mathbf{v}_1[k])+\mathbf{P}^{\top}(-\mathbf{b}+\mathbf{v}_2[k]))
  \end{align}
  \begin{align}\label{fin3}
    \mathbf{v}_1[k+1]=\mathbf{v}_1[k]+\mathbf{M}\mathrm{vec}(\mathbf{D}[k+1]+\mathbf{Q}[k+1])  
  \end{align}
  \begin{align}\label{fin4}
         \mathbf{v}_2[k+1]=\mathbf{v}_2[k]+\mathbf{P}\mathrm{vec}(\mathbf{D}[k+1])-\mathbf{b}
  \end{align}
  \begin{align}\label{fin5}
       \mathbf{v}_3[k+1]=\mathbf{v}_3[k]+\mathbf{R}\mathrm{vec}(\mathbf{Q}[k+1])
  \end{align}
  The complete procedure is summarized in Algorithm 1,
  After a number of iterations ($I_\text{max}$), we find $\mathbf{S}$ based on \eqref{shift}.
  \begin{algorithm}[h!]
\caption{Proposed ADMM-based solver}
\label{my_algorithm}
\begin{algorithmic}[1]
\Require  $ I_\text{MAX}, \mathbf{U}$.
%\ENSURE $\mathbf{S}$ is feasible, cost function is minimized
	\For{$i = 1$ to $I_\text{MAX}$}
\State{initialize $\mathbf{D}, \mathbf{v}_1,\cdots, \mathbf{v}_2$, $\mathbf{v}_3$ arbitrarily}
\State{update $\mathbf{Q}$ based on \eqref{fin1}}
\State{update $\mathbf{D}$ based on \eqref{fin2}}
\State{update $\mathbf{v}_1$ based on \eqref{fin3}}
\State{update $\mathbf{v}_2$ based on \eqref{fin4}}
\State{update $\mathbf{v}_3$ based on \eqref{fin5}}
\State \Return	 $\mathbf{D}, \mathbf{Q}, \mathbf{v}_1,\cdots, \mathbf{v}_2$, $\mathbf{v}_3$
\EndFor
\State{Obtain $\mathbf{S}=\mathbf{W}(\mathbf{D}+\mathbf{Q})\mathbf{W}^{\top}$}
\end{algorithmic}
\end{algorithm}
\noindent{where} $I_{\text{max}}$ denotes the maximum number of ADMM iterations.  
  \section{Numerical Results} 
This section describes numerical experiments that validate the performance of the proposed algorithms by averaging the results over 100 different random networks of $N$ nodes. The subspace matrix $\mathbf{U}$ is obtained by orthonormalizing an $N \times {r}$ matrix with i.i.d. standard Gaussian entries. Random signals $\boldsymbol{\alpha}$ and noise signals $\mathbf{v}$ were drawn from a normal distribution with zero mean and unit variance to generate input signals $\mathbf{z}$ such that $\mathbf{z}=\beta\sqrt{(N/r)}\mathbf{U}_{\parallel}\boldsymbol{\alpha}+\mathbf{v}=\boldsymbol{\xi}+\mathbf{v}$ where $\beta$ is Signal-to-Noise ratio (SNR). The graph  topology is generated through the Erdos-Renyi model~\cite{kolaczyck2009}, where the presence of each directed edge is an i.i.d. Bernoulli random variable.

The considered performance metrics are the Normalized Mean Projection Error (NMPE) and the Normalized Mean Square Error (NMSE), which are given by: $ \text{NMPE}(\mathbf{H}_{l})\triangleq\frac{
    \mathbb{E}_{\mathbf{z}}\left[||(\mathbf P -\mathbf H_l)\mathbf z||_2^2
      \right]
  }{
    \mathbb{E}_{\mathbf{z}}\left[||\mathbf P\mathbf z||_2^2\right]
  }$ and $\text{NMSE}({\hat{\boldsymbol{\xi}}}_{k})=
    \mathbb{E}\left[||\boldsymbol {\xi}-\hat{\boldsymbol{\xi}} ||_2^2
      \right]
  /
    \mathbb{E}\left[||\boldsymbol{\xi} ||_2^2\right] $, respectively. The expectation is taken over $\mathcal{G}$, $\mathbf{U}_{\parallel}, \alpha, \mathbf{v}$.
    
    The proposed method is compared with the other typical choices for the graph shift operator in previous works, such as the Laplacian matrix or the Adjacency matrix. Moreover, we also compare with a direct Least-Square method to find the asymmetric graph shift operator, by minimizing: $||\mathrm{vec}(\mathbf{H})-\mathbf{G}\mathrm{vec}(\mathbf{S})||_2^2$, where $\mathbf{G}$ is a matrix that satisfies the topology constraints, i.e. $\mathbf{S}_{n,n'}=0$ $\forall{(v_n,v_{n'})}\not\in{\mathcal{H}}$ via $\mathbf{G}\mathrm{vec}(\mathbf{S})$.   For all the methods, the filter coefficients are found by solving directly the equation  $\mathbf{H}:=\sum_{l=0}^{L}c_l\mathbf{S}^{l}$. Besides, to alleviate problems associated with finite-precision arithmetic, each node uses a different set of filter coefficients~\cite{segarra2017operators}. It can be easily shown that all the results of the paper carry over also when each node uses a different set of filter coefficients. %Moreover, for the proposed method, the identity matrix is chosen as $\mathbf{E}_\perp$ and $\mathbf{E}_\parallel$. 
    
\begin{figure}[h]
\centering
\includegraphics[width=0.7\textwidth]{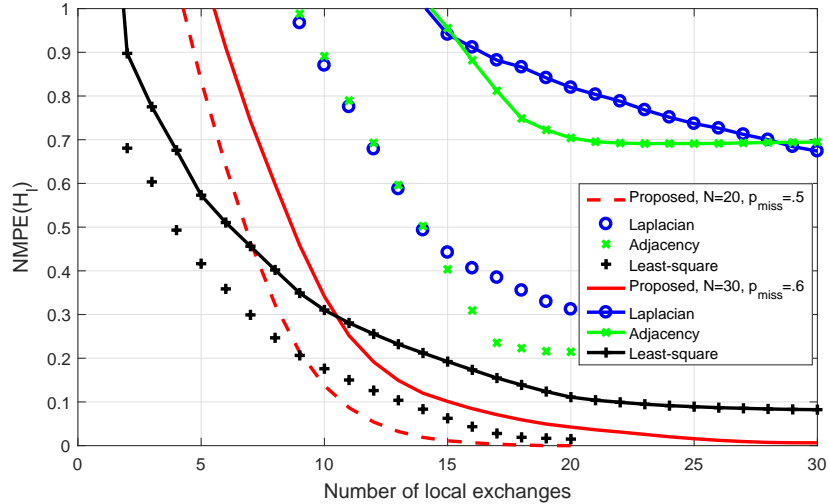}
\caption{NMPE as a function of number of local exchanges ($r=3,\epsilon=1, \beta=5$, $\rho=0.1$, $I_{\text{max}}=1000$) }
\label{NMPE}
\end{figure}

    Figure \eqref{NMPE} depicts NMPE versus the number of local exchanges for different scenarios. It can be observed that the proposed method outperforms all the other methods, and achieves the exact projection after a small finite number of iterations, thus showing that it converges faster than the other methods.
    
\begin{figure}[h]
\centering
\includegraphics[width=0.6\textwidth]{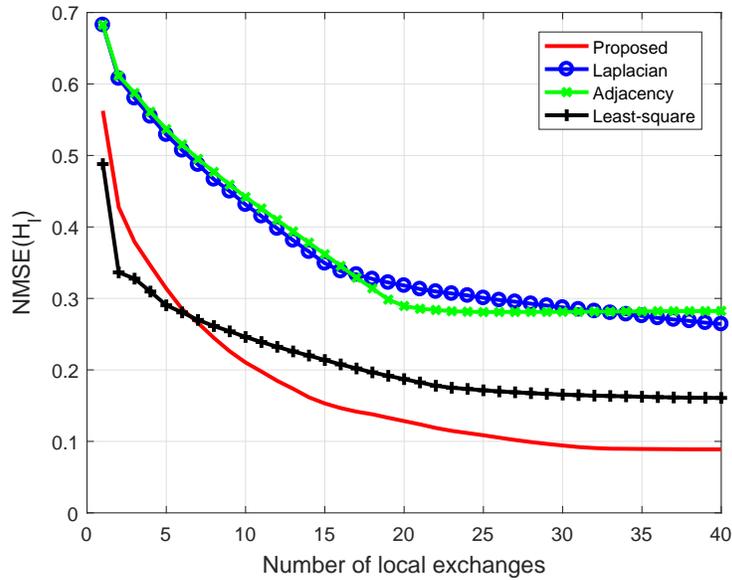}
\caption{NMSE as a function of number of local exchanges ($N=40, r=4, p_{miss}=.7, \epsilon=1,\beta=5$, $\rho=0.1$, $I_{\text{max}}=1000$)}
\label{NMSE}
\end{figure}

    Figure \eqref{NMSE} shows the NMSE versus the number of local exchanges for sparser and larger networks, comparing our method with the other choices of graph shift operator. Our proposed method obtains lower NMSE in comparison with all the other methods.    
    \section{Conclusion}
    This paper proposes an algorithm to design asymmetric graph shift operator to compute decentralized subspace projection. The proposed method is also capable to design other linear transformations. The results show that the proposed method obtain the projection matrix exactly after a finite number of iterations.   
\bibliographystyle{ieeetr}
\bibliography{refs}
\end{document}